\documentstyle[12pt]{article}
\def\bib{\bibitem}
\def\be{\begin{equation}}
\def\ee{\end{equation}}
\def\barr{\begin{array}}
\def\earr{\end{array}}
\def\dis{\displaystyle}

\def\etal{ {\em et al.}}
\def\ie{ {\em i.e.}}
\def\viz{ {\em viz.}}

\def\lsim{\:\raisebox{-0.5ex}{$\stackrel{\textstyle<}{\sim}$}\:}
\def\gsim{\:\raisebox{-0.5ex}{$\stackrel{\textstyle>}{\sim}$}\:}
\def\gev{\: {\rm GeV} }

\def\ra{\rightarrow}
\def\mand{\qquad {\rm and} \qquad}

\def\Ch{\widetilde{\chi_1}}

\def\Ne{\widetilde{\chi^0_1}}
\def\neii{\widetilde{\chi^0_i}}
\def\nejj{\widetilde{\chi^0_j}}
\def\ntwo{\widetilde{\chi^0_2}}
\def\nonetwo{\widetilde{\chi^0}_{\hspace{-0.4em}1,2}}
\def\rbs{\delta R_b^{\rm SUSY}}
\def\etsl{$E_T \hspace{-1.15em}/\;$}

\def\ib#1,#2,#3{       {\it ibid.\/ }{\bf #1} (19#2) #3}
\def\ap#1,#2,#3{       {\it Ann.~Phys.~(NY)\/ }{\bf #1} (19#2) #3}
\def\ijmp#1,#2,#3{     {\it Int.~J.~Mod.~Phys.\/ } {\bf A#1} (19#2) #3}
\def\mpl#1,#2,#3 {     {\it Mod.~Phys.~Lett.\/ } {\bf A#1} (19#2) #3}
\def\np#1,#2,#3{       {\it Nucl.~Phys.\/ }{\bf B#1} (19#2) #3}
\def\npps#1,#2,#3{     {\it Nucl.~Phys.~B (Proc.~Suppl.)\/ }{\bf B#1}
                             (19#2) #3}
\def\plb#1,#2,#3{      {\it Phys.~Lett.\/ }{\bf B#1} #2 (19#3)}
\def\pr#1,#2,#3{       {\it Phys.~Rev.\/ }{\bf #1} #2 (19#3)}
\def\prd#1,#2,#3{       {\it Phys.~Rev.\/ }{\bf D#1} #2 (19#3)}
\def\prep#1,#2,#3{     {\it Phys.~Rep.\/ }{\bf #1} #2 (19#3)}
\def\prl#1,#2,#3{      {\it Phys.~Rev.~Lett.\/ }{\bf #1} #2 (19#3)}
\def\pro#1,#2,#3{      {\it Prog.~Theor.~Phys.\/ }{\bf #1} #2 (19#3)}
\def\rmp#1,#2,#3{      {\it Rev.~Mod.~Phys.\/ }{\bf #1} #2 (19#3)}
\def\sp#1,#2,#3{       {\it Sov.~Phys.-Usp.\/ }{\bf #1} #2 (19#3)}
\def\zpc#1,#2,#3{      {\it Zeit.~f\"ur Physik\/ }{\bf C#1} #2 (19#3)}
\begin{document}
\setcounter{page}{0}
\renewcommand{\thefootnote}{\fnsymbol{footnote}}
\thispagestyle{empty}
\vspace*{-1in}
\begin{flushright}
CERN-TH/96-203\\[1.5ex]
TIFR/TH/96-41\\[1.5ex]
{\large \tt hep-ph/9608264} \\
\end{flushright}

\vskip 35pt
\begin{center}
{\Large \bf An $R$--Parity Breaking SUSY Solution to  \\[1.0ex]
            the $R_b$ and ALEPH Anomalies}

\vspace{10mm}
{\large Debajyoti Choudhury$^{(1),}$\footnote{debchou@mppmu.mpg.de}
and D.P. Roy$^{(2,3),}$\footnote{dproy@theory.tifr.res.in}
}\\[1.5ex]

$^{(1)}${\em Max--Planck--Institut f\"ur Physik,
              Werner--Heisenberg--Institut,\\
              F\"ohringer Ring 6, 80805 M\"unchen,  Germany.}\\[1.0ex]
$^{(2)}${\em Theory Division, CERN, CH--1211 Geneva 23,  
        Switzerland.}\\[1.0ex]
$^{(3)}${\em Theoretical Physics Group, 
             Tata Institute of Fundamental Research,\\ 
Homi Bhabha Road, Bombay 400 005, India.}\\
\vspace{50pt}
{\bf ABSTRACT}
\end{center}

\begin{quotation}
We discuss an optimal $R$-parity breaking SUSY solution to the 
$R_b$ excess as well as the ALEPH 4-jet anomaly. The latter arises 
from the pair production of stop via chargino decay at LEP1.5, 
followed by its  $R$--violating decay into a light quark pair. The model 
satisfies top quark and $Z$--boson decay constraints along with gaugino 
mass unification.
\end{quotation}

\vspace{90pt}
\noindent
CERN-TH/96-203\\[1.5ex]
August, 1996

\vfill
\newpage
\setcounter{footnote}{0}
\renewcommand{\thefootnote}{\arabic{footnote}}

\setcounter{page}{1}
\pagestyle{plain}
\advance \parskip by 10pt

Two of the intriguing results from LEP which have attracted a good deal 
of theoretical interest are the $R_b$ and the ALEPH 4-jet anomalies. 
The first anomaly refers to the LEP1 value of 
$R_b (\equiv \Gamma_Z^{b \bar b} /\Gamma_Z^{\rm had})$ being 
$\sim 2 \sigma$ larger than the SM prediction~\cite{lepewwg,blondel}.
The second refers to the anomalous 4-jet events recently 
reported~\cite{aleph4j} by the ALEPH experiment at LEP1.5, each of 
which seems to consist of dijet pairs with a common invariant mass 
$~\sim 55 \gev$. 

It is now widely recognized~\cite{rb_old,phc_stp_96,tifr5} that 
the minimal supersymmetric standard model (MSSM) offers a viable 
solution to the $R_b$ anomaly in the low $\tan \beta$ region if 
one assumes a relatively light top squark ($\tilde t_1$) and 
chargino ($\Ch$). The  assumption of a $R$--parity 
violating Yukawa interaction term would invalidate the canonical 
missing $E_T$ (\etsl) signature for superparticle production
and thus the LEP1.5 limit on chargino mass~\cite{LEP1.5},
\be
      m_{\Ch} > 65 \gev \ .
              \label{ch_mass_lim}
\ee
In this case, pair production of charginos at LEP1.5 can offer a possible 
explanation for the ALEPH 4-jet events as was  recently suggested in 
refs.~\cite{ggr,herbi,phc_dc_stp}. In particular, Chankowski 
\etal\cite{phc_dc_stp} have discussed a variety of such $R$--parity 
breaking SUSY solutions to the ALEPH anomaly, which can also account 
for the $R_b$ excess. The purpose of this note is to focus on what appears 
to be an optimal $R$--parity breaking 
SUSY solution to the $R_b$ and ALEPH anomalies, 
within the constraints of top quark and $Z$--boson decays as well as 
that of gaugino mass unification. We shall see below that 
it can quantitatively account for the essential features of the 
ALEPH 4-jet events as well as for the $R_b$ anomaly.

Explicit breaking of $R$-parity introduces additional Yukawa terms
in the superpotential~\cite{rpar}
\be
W_{\not R}
  = \frac{1}{2} \lambda_{ijk} L_i L_j E^c_k
  + \lambda'_{ijk} L_i Q_j D^c_k
  + \frac{1}{2} \lambda''_{ijk} U^c_i D^c_j D^c_k \ ,
                                       \label{superpot}
\ee
where $Q,\ L$ ($U,\ D,\ E$) denote the quark and lepton 
doublet (singlet) superfields and the subscripts denote the 
generation. Symmetry considerations imply that 
$\lambda_{ijk} = -\lambda_{jik}$ and
$\lambda^{\prime\prime}_{ijk} = -\lambda^{\prime\prime}_{ikj}$.
Proton stability demands that all products of the 
form ($\lambda' \lambda''$) be vanishingly small and this, 
conventionally, is ensured by stipulating that either the baryon 
number violating 
couplings ($\lambda''$) or the lepton number violating 
couplings ($\lambda, \lambda'$) are non-zero but not both. 

Let us briefly discuss the various $R$--parity breaking SUSY 
scenarios that have been suggested as explanations of the 4-jet 
excess. The first of these~\cite{bkp}, which, in fact, predates 
the anomaly, assumes the lightest superparticle (LSP) to be a 
sneutrino instead of the neutralino. Pair production of 
sneutrinos ($e^+ e^- \ra \tilde{\nu}^\ast \tilde{\nu}$) followed
by their diquark decays---through one of the $\lambda'$ 
couplings---can then lead to a 4-jet final state. An adequate production
cross section can be obtained for $\tilde \nu_e$ (via a light, and
gaugino-dominated, chargino exchange), 
but the other event characteristics have not been
analysed so far. The second scenario~\cite{peccei} suggests pair 
production of left-handed $b$ squarks, followed by their diquark decays
through a $\lambda''$ coupling. However, a light $\tilde b_L$ in the 
required mass range ($\sim 55 \gev$) seems to be disfavoured by the
precision measurements of electroweak observables~\cite{phc_dc_stp}. 
More recently, there have been two suggestions based on pair production 
of charginos with mass $\sim 55 \gev$~\cite{ggr,herbi}. One of 
them~\cite{ggr} assumes the decay sequence
\be
\Ch \ra W^* \: \Ne \ , \qquad W^* \ra \bar{q} q' \ ,\qquad 
  \Ne \; \raisebox{-0.5ex}{$\stackrel{\textstyle \lambda''}
                     {\longrightarrow}$} \; q_1 \: q_2 \: q_3 \ ,
         \label{ggr_decay}
\ee
where the star denotes an off-shell $W$ boson. Although this is the most
natural scenario in terms of the MSSM mass spectrum, it seems to be
disfavoured on several counts. It shows much broader distributions
in the difference of the dijet masses as well as their sum than 
the ALEPH data~\cite{aleph4j}. Moreover, the leptonic decay of one 
of the $W^*$s would imply roughly as many anomalous events with an isolated 
lepton as without it, which could not have been missed. Finally, one 
would have a significantly large fraction of the events with more 
than 4 jets if one applies the ALEPH jet algorithm. The other 
suggestion~\cite{herbi} assumes the chargino decay
\be
   \Ch \; \raisebox{-0.5ex}{$\stackrel{\textstyle \lambda'}
                     {\longrightarrow}$} \; \tau \bar{q} q' \ 
               \label{herbi_decay}
\ee
to dominate over its $R$--conserving decay into the LSP. In this case,
the outgoing $\tau$ would be too hard to have been missed, unless one 
assumes the exchanged sneutrino mass to be very close to that of the 
chargino. It may be noted here that none of the above scenarios address
the issue of the $R_b$ anomaly, to which we  now turn. 

We start by considering the standard $R$-conserving MSSM solution 
to the $R_b$ anomaly, within the constraints of top quark and $Z$ 
boson decays~\cite{phc_stp_96,tifr5}. Gaugino mass unification shall
be assumed all along as it is very closely related to the 
successful MSSM prediction for the unification of the 
$SU(3) \otimes SU(2) \otimes U(1)$ gauge couplings. Thus, the 
masses of the corresponding gauginos $\tilde g$, $\tilde W$ and 
$\tilde B$ are related via
\be \displaystyle
   M_3 = \frac{\alpha_s}{\alpha} \sin^2 \theta_W M_2 \simeq 3.5 M_2\ 
      \mand
   M_1 = \frac{5}{3} \tan^2 \theta_W M_2 \simeq 0.5 M_2\ .
               \label{mass_unif}
\ee
The physical gluino mass is related to the running mass through the 
QCD correction factor~\cite{gluino_mass_qcd}
\be \displaystyle
     m_{\tilde g} = \left( 1 + 4.2 \frac{\alpha_s}{\pi} \right) M_3
                  \simeq 1.15 M_3 \simeq 4 M_2 \ .
     \label{gluino_mass}
\ee
Thus, a single gaugino mass ($M_2$) along with the higgsino mass 
parameter ($\mu$) and the ratio of the two Higgs vacuum expectation
values ($\tan \beta$), determine the gluino mass as well as the 
masses and compositions of the two chargino and four neutralino 
states~\cite{haber_kane} \ie,
\be
 \barr{rcl}
   \widetilde{\chi}_{iL}^\pm & = &  V_{i1} \widetilde{W}_L^\pm 
                                  + V_{i2} \widetilde{H}_L^\pm \ ,
   \qquad
   \widetilde{\chi}_{iR}^\pm  =   U_{i1} \widetilde{W}_R^\pm 
                                  + U_{i2} \widetilde{H}_R^\pm \ ,
   \\[1.5ex]
      \neii & = &  
          N_{i1} \widetilde{B} + N_{i2} \widetilde{W_3}
        + N_{i3} \widetilde{H_1^0} + N_{i4} \widetilde{H_2^0} \ .
\earr
         \label{gaugino_mixing}
\ee

In the scalar sector, the large Yukawa term for the top results in 
a mass hierarchy and thus the lighter (and predominantly right-handed) 
stop,
\be
    \tilde t_1 \equiv \cos \theta_{\tilde t} \: \tilde t_R
                  +   \sin \theta_{\tilde t} \: \tilde t_L \ ,
       \label{stop_eigen}
\ee
is expected to be significantly lighter than the other squarks. We shall
be primarily interested in this stop.

In the low $\tan \beta$ region of our interest, the SUSY contributions
to $Z \ra b \bar{b}$ arise from the triangle graphs involving 
$\tilde \chi_i \tilde \chi_j \tilde t_k$ and  
$\tilde t_i \tilde t_j \tilde \chi_k$ exchanges as well as the 
$\tilde t_i \tilde \chi_j$ loop insertions in the $b$ and $\bar{b}$ 
legs~\cite{phc_stp_96,tifr5,boul_fin}. The $b$ vertices are dominated by 
the $b_L \tilde t_1 \tilde \chi_i$ Yukawa coupling
\be \dis
       \Lambda^L_{1 i} \simeq - \: \frac{m_t V_{i2} \cos \theta_{\tilde t} }
                                 { \sqrt{2} m_W \sin \beta}
    \label{b_st_chi vertex}
\ee
which favours large $V_{12}$, \ie, the higgsino--dominated region 
($|\mu| \ll M_2$). On the other hand, the $Z \tilde \chi_i \tilde \chi_j$
couplings
\be \dis
       O^L_{i j} = - \frac{1}{2} \left( \cos 2 \theta_W \delta_{i j} 
                                        + U_{i1} U_{j 1} \right) 
                  \mand 
       O^R_{i j} = - \frac{1}{2} \left( \cos 2 \theta_W \delta_{i j} 
                                        + V_{i1} V_{j 1} \right) \ ,
    \label{Z_chi_chi vertex}
\ee
favour large $U_{11}$ and $V_{11}$, \ie, the gaugino--dominated region
($|\mu| \gg M_2$). (The corresponding $Z \tilde t_1 \tilde t_1 $ coupling
is suppressed by the $U(1)$ coupling 
factor $\sin^2 \theta_W$.) The need for sizeable 
$b$ as well as $Z$ couplings then implies that the largest SUSY 
contribution to $R_b$ ($\rbs$) occurs for the mixed 
region ($|\mu | \sim M_2 $)----corresponding to a 
$\tilde \gamma$--dominated LSP---rather 
than for the higgsino--dominated 
region~\cite{phc_stp_96,tifr5}. 
Moreover, it seems to favour negative $\mu$ 
over the positive $\mu$ region~\cite{phc_stp_96}. We shall,
therefore, restrict ourselves to the former. We shall consider
\be
      \rbs \sim 0.0020 {\mbox{ \rm --- }} 0.0025
            \label{delta R_b}
\ee
to be a viable solution to the $R_b$ anomaly. It would exactly account 
for the discrepancy between the current experimental value~\cite{blondel}
of $R_b^{\rm exp} = 0.2178 \pm 0.0011$ and the standard model value 
(for $m_t = 175 \gev$) of $R_b^{\rm SM} = 0.2156$, as well as close 
the gap between the $\alpha_s (m_Z^2)$ estimates from LEP1 and from 
deep inelastic scattering.

Low masses for $\tilde t_1$ and $\Ch$, required for a suitably 
large $\rbs$, may, however, result in significant new decay 
channels for the top quark, \viz, 
 $t \ra \tilde t_1 \nonetwo$.
The decay amplitudes are dominated by 
the $\bar{t}_L \neii \tilde t_{1R}$ 
and $\bar{t}_R \neii \tilde t_{1L}$ Yukawa 
couplings~\cite{phc_stp_96,tifr5,bdggt} :
\be \dis
   C_i^{L(R)} \simeq \frac{m_t N_{i 4} } 
                          { m_W \sin \beta} 
                     \: \cos \theta_{\tilde t} \; 
                        (\sin \theta_{\tilde t})         \ ,
       \label{stop yukawa}
\ee
with $i = 1, 2$. The higgsino--dominated region corresponds to large 
$\tilde H_2^0$ components in $\nonetwo$, and hence 
a large SUSY Branching Ratio ($B_S$) for top decay. This quantity is
relatively small in the mixed region since only $\ntwo$
has a large $\tilde H_2^0$ component while 
$\Ne \simeq \tilde \gamma$. Thus the upper limit 
on $B_S$ from the CDF top decay data~\cite{CDFtop} favours the mixed 
region as well. We shall take a rather lenient value for this 
limit~\cite{phc_stp_96,tifr5}:
\be 
         B_S < 0.4 \ .
                 \label{B_S limit}
\ee

Recently, a systematic scan of the MSSM parameter space was carried 
out~\cite{phc_stp_96} to obtain the best SUSY contribution to $R_b$
within the constraint of top quark decay. The optimal value of
$\rbs$ is obtained at small negative value of the stop mixing angle
($\theta_{\tilde t} \approx -15^\circ$) and small stop mass
\be
    m_{\tilde t_1} \sim 50 {\mbox{ \rm --- }}60 \gev \ ,
             \label{m_st for rb}
\ee
which lies in between the LEP1 limit 
($m_{\tilde t_1} > 45 \gev$)~\cite{PDG} and the D0 excluded region 
($m_{\tilde t_1} \ne 65$--85 GeV)~\cite{D0stop}. LEP1.5 imposes 
no additional bound on $m_{\tilde t_1}$ as the pair-production 
cross section for the right-handed stop is small. 

\begin{table}[h]
\begin{center}

\begin{tabular}{|c|c|c|c|c|c|c|c|c|c|}
\hline
&&&&&&&&&\\
& ($M_2, \mu$) & $\tan \beta$  
    & $ \Gamma(Z \ra \neii \: \nejj ) $
    & $ m_{\Ch} $   &   $ m_{\Ne} $ & $ m_{\ntwo} $
    & $ m_{\tilde t_1}$ & $\rbs $ & $B_S$ \\[1.5ex]
\hline
$A$ & (150, $-40$) & 1.4 & ---   & 67 & 39 & 70 & 60 & 0.0014 & 0.51 \\
   & (150, $-30$) & 1.4 & 3 MeV & 58 & 29 & 71 & 50 & 0.0019 & 0.54 \\
&&&&&&&&&\\
\hline
$B$ & (60, $-60$) & 1.4 & ---   & 86 & 35 & 57 & 60 & 0.0019 & 0.40 \\
&&&&&&&&&\\
\hline
$C$ & (40, $-70$) & 1.4 & ---   & 76 & 24 & 64 & 60 & 0.0021 & 0.30 \\
   &           & 2.0 & 0.6 MeV  & 64 & 24 & 52 & 55 & 0.0021 & 0.26 \\
   &           & 2.6 & 2.2 MeV  & 56 & 23 & 45 & 50 & 0.0024 & 0.34 \\
\hline
\end{tabular}
\end{center}
   \caption{\em SUSY contributions to $R_b$ ($\rbs$) and to the
        top BR ($B_S$) 
      are shown, along with the light chargino and neutralino masses,
      for three representative points in the ($M_2, \mu$) plane.
      (All masses are in GeV and $\theta_{\tilde t} = - 15^\circ$.) For
      each set, the top row corresponds to the allowed MSSM parameter 
      space in the $R$-conserving scenario.}
\end{table}

In Table 1 we display the phenomenological consequences for 
three representative points in the ($M_2, \mu$) plane :
\be
 A\: : \ (150, -40) \gev, \qquad B \: : \ (60, -60) \gev \mand
      C \: : \ (40, -70) \gev,
      \label{repr points}
\ee
belonging to the higgsino--dominated ($A$) and mixed regions ($B,C$).
The points are chosen close to the LEP limit so as to give the best 
values of $\rbs$ in the respective regions. For the higgsino--dominated 
point ($A$), $m_{\Ch}$ is close to its LEP1.5 
limit (\ref{ch_mass_lim}). Still $\rbs$ is smaller than the required 
value (\ref{delta R_b}), while the contribution to top BR exceeds 
the limit (\ref{B_S limit}). On the other hand, the mixed region 
points ($B,C$) are seen to give viable values of $\rbs$, while satisfying
the $B_S$ limit. 
Note that, in this region, the chargino mass is safely above the LEP1.5
limit (\ref{ch_mass_lim}); the most important constraint is rather 
set by the requirement that
\be
     m_{\Ne} + m_{\ntwo} \gsim m_Z
           \label{sum_of m_neut}
\ee
which follows from the stringent LEP1 bound
\be 
        BR (Z \ra \Ne \: \ntwo ) 
                   < 5 \times 10^{-5}
      \label{Z_n1n2}
\ee
deduced from the negative search for acoplanar jets using the 
\etsl\ signature of $\Ne$~\cite{PDG}. The 
best values for $\rbs$ and $B_S$ are  obtained for point ($C$). 
However, it corresponds to a gluino mass of 160~GeV, barely above 
the Tevatron lower limit~\cite{D0gluino}. 

We now turn to effects of $R$-parity breaking via the $\lambda''$ 
couplings. As the LSP ($\Ne$) now undergoes hadronic decay, the 
\etsl\ signature is no longer applicable. Thus, the LEP1.5 
bound (\ref{ch_mass_lim}) is inoperative. Moreover, the LEP1 
bound (\ref{Z_n1n2}) on $Z$ decay into neutralinos 
is now replaced by 
\be 
     \sum \Gamma(Z \ra \neii \nejj )
             < 3 \: {\rm MeV}
    \label{Z_ni_nj}
\ee
corresponding to the $1 \sigma$ error in $\Gamma_Z^{\rm had}$. 
This is weaker than (\ref{Z_n1n2}) by more than an order of magnitude. 
Nonetheless, as we shall see below, 
it acts as a strong constraint on efforts to reduce 
the chargino mass below 65~GeV in the mixed region. 

The goal then is to have $m_{\Ch} < 65 \gev$, while satisfying 
(\ref{Z_ni_nj}) so that chargino pair production at LEP1.5 can 
be a source for the anomalous 4-jet events. In the higgsino--dominated 
region this is achieved most easily by decreasing $|\mu|$, while in 
the mixed region it can be achieved only by increasing $\tan \beta$. 
The former also has the advantage of increasing $\rbs$ simultaneously. 
Table 1 shows that it is possible to go down to $m_{\Ch} = 58 \gev$
within the constraint (\ref{Z_ni_nj}) in the higgsino--dominated region. 
The stop mass can then be reduced to 50~GeV so as to allow the 
two body decay mode
\be
    \Ch \ra \tilde t_1 b \ ,
                    \label{ch_2body}
\ee
which shall be assumed later on. Reducing the chargino and the stop masses 
has the effect of increasing $\rbs$ to the respectable value of 0.0019. 
Unfortunately, $B_S$ is untenably large.

For  point $B$  ($|\mu| = M_2$), it is not possible to drive down 
$m_{\Ch}$ below $\sim 65 \gev$ 
by either of the above methods, while still 
satisfying (\ref{Z_ni_nj}). However, for the mixed region point 
$C$---which offers the best value for $\rbs$---it is possible to achieve
this by increasing $\tan \beta$. As Table 1 shows, one can go down to
$m_{\Ch} = 56 \gev$, within the above constraints,
by increasing $\tan \beta$ to 2.6. Decreasing
the stop mass to 50~GeV ensures the two-body decay (\ref{ch_2body}) with
a soft $b$. Note that the large value for $\rbs$ is obtained 
within the $B_S$ constraint (\ref{B_S limit}). Also, the value of 
$\tan \beta$ is now more reasonable. Furthermore, the gluino mass 
limit from the Tevatron is no longer applicable.

We shall quantitatively pursue the $R$--parity breaking SUSY scenario 
summarised in the last row of Table 1 as a possible solution to the 
anomalous 4-jet events. The pair production of charginos, followed by 
their two-body decays (\ref{ch_2body}), results in 
$\tilde t_1 \tilde t_1^\ast$ along with a soft $\bar b b $ pair. For
\be
      m_{\Ch} = 56 \gev \mand m_{\tilde t_1} = 50 \gev \ ,
    \label{masses}
\ee
the $b$ momenta are always less than 6~GeV and so the $b$s are 
expected to largely 
miss the lifetime tag. Furthermore, leptons from $b$-decay do not 
survive isolation cuts. 
The stop can decay either directly
\be 
    \tilde t_1^\ast \; \raisebox{-0.5ex}{$\stackrel{\textstyle \lambda''}
                     {\longrightarrow}$} \; d \: s \ ,
        \label{st_ds}
\ee
or through the $R$--conserving loop process
\be 
    \tilde t_1 \ra c \Ne \ ,
        \label{st_cn}
\ee
followed by
\be 
    \Ne \; \raisebox{-0.5ex}{$\stackrel{\textstyle \lambda''}
                     {\longrightarrow}$} \; u\: d\: s \ , c\: d\: s 
                          \ .
        \label{neut_decay}
\ee
We do not consider stop decay modes with $b$ quarks in the final state
since these will lead 
to a large number of $b$-tags in conflict with the ALEPH 
results~\cite{aleph4j}. The loop decay (\ref{st_cn}) is a third-order 
electroweak process and hence has a very small width~\cite{hik_kob}. 
Consequently, the $R$-violating decay (\ref{st_ds}) dominates over a very
large range of the Yukawa coupling 
\be
     \lambda''_{tds} \gsim 5 \times 10^{-5} \ .
         \label{tds_limit}
\ee
On the other hand, one 
requires \cite{goity_sher}
\be
     \lambda''_{uds,cds} \gsim 5 \times 10^{-3} 
         \label{uds_limit}
\ee
for the LSP decay to occur within 1 cm~\cite{aleph_neutdecay}. Note that 
the direct decay (\ref{st_ds}) dominates as long as $\lambda''_{tds}$
is larger than the relatively modest limit of (\ref{tds_limit}) 
{\em irrespective} of the other Yukawa couplings. Thus the direct 
decay is at least as 
natural\cite{lsp}
as the alternative route of (\ref{st_cn} \& \ref{neut_decay}). We shall 
see below that the former can quantitatively account for the ALEPH 
events, while the latter cannot.

\begin{figure}[t]
\vskip 4.8in\relax\noindent\hskip -0.6in\relax{\includegraphics{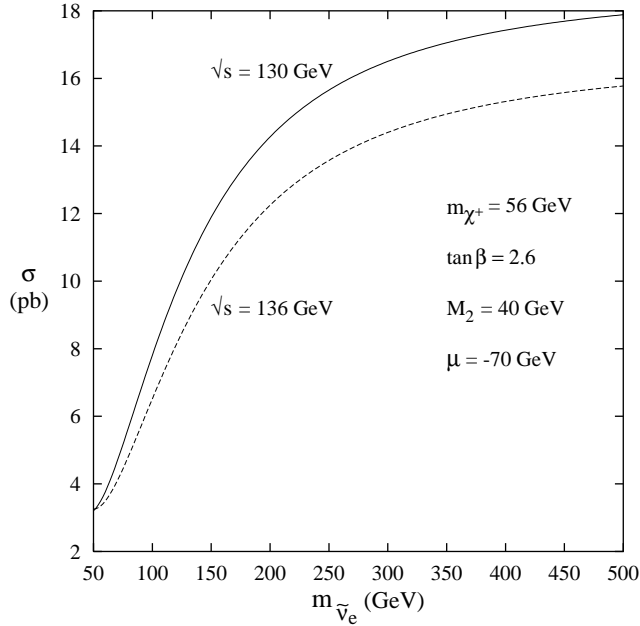}}
\vspace{-20ex}
\caption{ {\em Pair production cross section for a 56 GeV chargino 
as a function of the sneutrino mass for the two different LEP1.5 
center-of-mass energies.} }
     \label{fig:cross_sec}
\end{figure}

Figure~\ref{fig:cross_sec} shows the chargino pair-production 
cross-section at the (LEP1.5) energies of 130 and 136 GeV. Since the 
interference between the $s$-channel ($\gamma/Z$) and the $t$-channel
($\tilde \nu_e$) is a destructive one, the cross section 
increases strongly with the sneutrino mass. For the rest of the 
analysis we have averaged the cross sections at the two energies assuming 
a sneutrino mass of 200~GeV.

We have studied chargino pair-production and subsequent decay via 
the stop (\ref{ch_2body}) using a parton level Monte Carlo program. 
Both the two-body (\ref{st_ds}) and the 
four-body (\ref{st_cn} \& \ref{neut_decay}) decay modes have been considered. 
To estimate the effect of jet energy resolution, we have compared 
the results with and without the suppression of soft partons (arising 
mainly from $b$-decays), having 
energy less than the ALEPH resolution error~\cite{aleph4j}
\be
    \sigma_E = ( 0.6 \sqrt{E (\gev)} + 0.6) \gev \ (1 + \cos^2 \theta)
                   \ .
            \label{ener_res}
\ee
We found the difference to be small. The results presented below 
correspond to the suppression of the soft partons having energy 
$< \sigma_E$.

\begin{figure}[h]
\vskip 4.8in\relax\noindent\hskip -0.6in\relax{\includegraphics{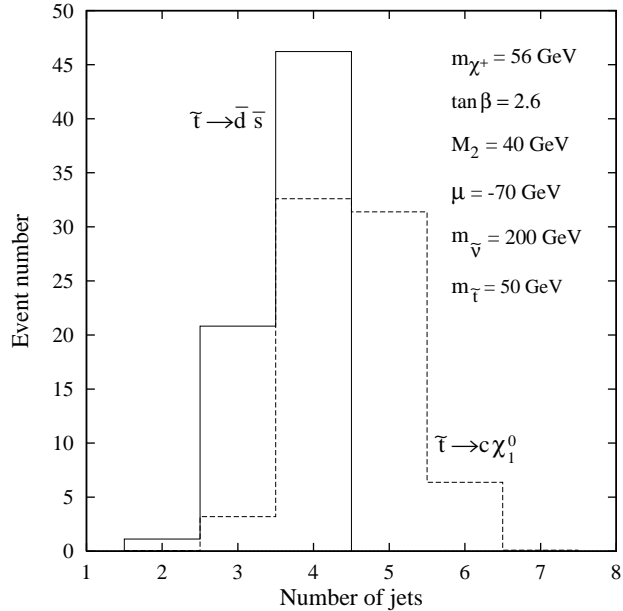}}
\vspace{-20ex}
\caption{ {\em Event distribution as a function of the number of jets 
    after the initial partons have been clustered with Durham/JADE 
    algorithms (\protect\ref{dur-cut}, \protect\ref{jade-cut}). 
    Both decay modes of the stop are shown.} }
     \label{fig:jet_distr}
\end{figure}
The parton jets are merged applying the Durham algorithm till~\cite{aleph4j}
\be
y_{\rm Dur} \equiv 2 \;{\rm min}(E_i^2, E_j^2) \ (1-\cos\theta_{ij})/ s
       > 0.008\ .
            \label{dur-cut}
\ee
Events merging into less than 4 jets are reclustered~\cite{aleph4j} with
the JADE algorithm till
\be
y_{\rm JADE} \equiv 2 E_i E_j \ (1-\cos\theta_{ij}) / s > 0.022 \ .
       \label{jade-cut}
\ee
Figure~\ref{fig:jet_distr} shows the  resulting distributions in 
the number of jets for the two decay modes in question. For the direct 
decay mode, the 4-jet sample dominates in agreement with 
ref.\cite{aleph4j}, while for the neutralino-mediated case the 
number of 5-jet events is uncomfortably large. In the 
former case, QCD radiation effects could result in a few 5-jet 
configurations as observed in \cite{aleph4j}. 

\begin{figure}[ht]
\vskip 4.8in\relax\noindent\hskip -1.85in
                          \relax{\includegraphics{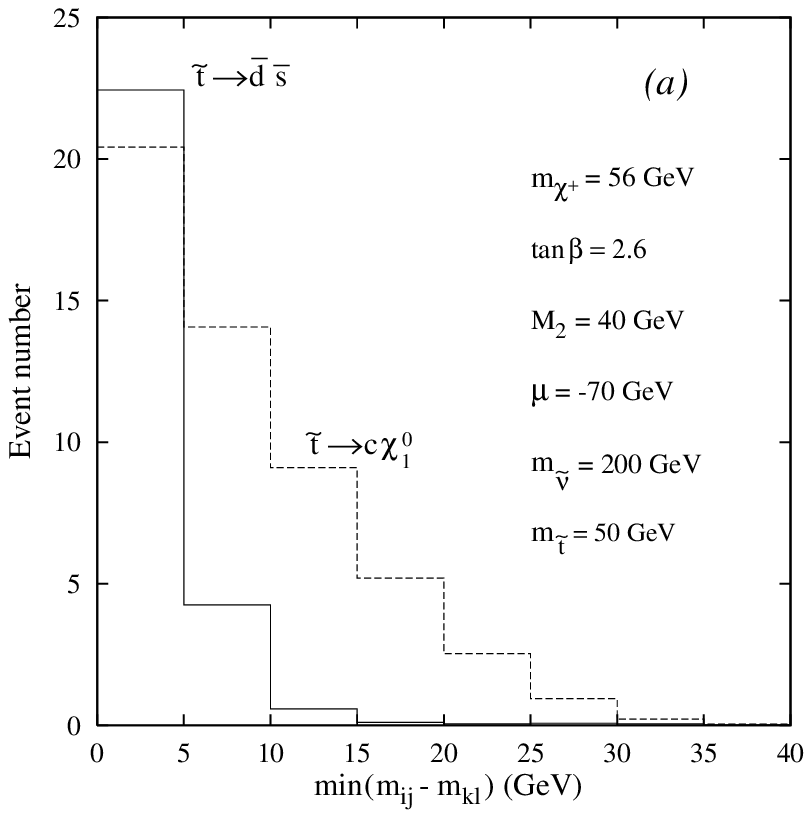}}
            \relax\noindent\hskip 3.20in
                          \relax{\includegraphics{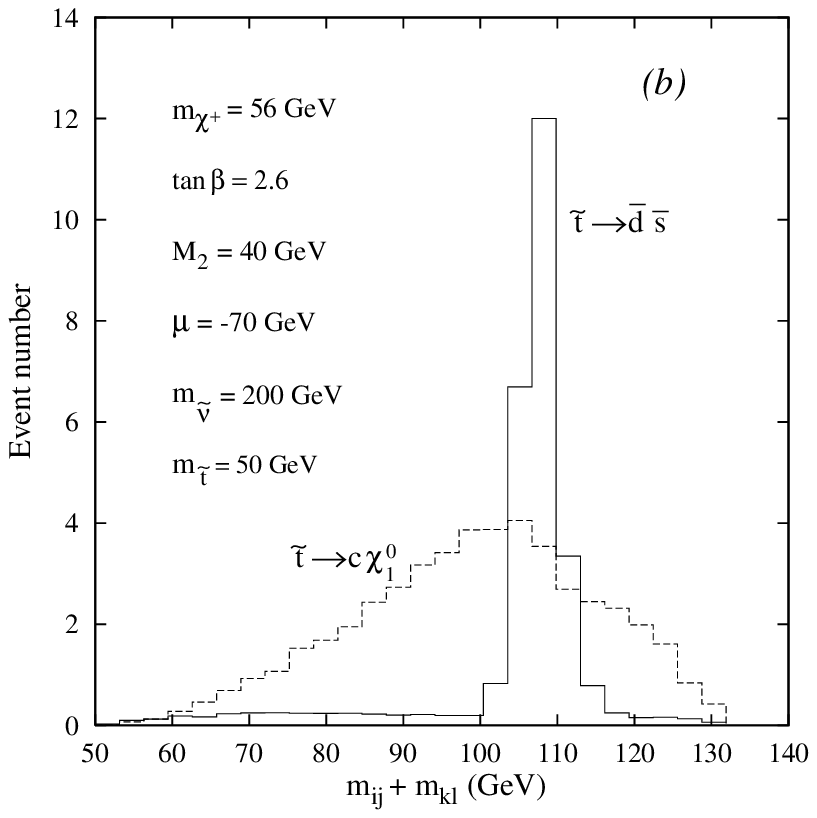}}
\vspace{-20ex}
\caption{ {\em Event distribution (after clustering down to 4 jets) 
    and rejecting events with smallest dijet invariant mass 
    $< 25 \gev$. {\em (a)} As a function of the minimum difference 
    of the dijet invariant masses. 
    {\em (b)} As a function of the sum of the dijet invariant 
    masses for the pair with the smallest mass difference. } }
     \label{fig:mass_distr}
\end{figure}

The 5-jet events of Fig.~\ref{fig:jet_distr} are then clustered down
to 4 by merging the two jets with the smallest invariant mass. 
Within this sample, events having the smallest dijet invariant mass 
$< 25 \gev$ are rejected~\cite{aleph4j}. This reduces the number of 
events by $\sim 40 \%$, which is compatible with ref.\cite{aleph4j}. 
Figure~\ref{fig:mass_distr}({\em a})
shows the distribution of the remaining 4-jet 
events in the minimum difference of the dijet invariant masses. At this
stage, both  distributions are in reasonable agreement with 
ref.\cite{aleph4j}. QCD radiation effects are expected to 
cause a marginal broadening of these distributions.

Figure~\ref{fig:mass_distr}({\em b}) shows the corresponding distributions 
in the sum of the dijet invariant masses (the pairing decided 
by the minimum difference). For the direct stop decay (\ref{st_ds}),
this distribution is sharply peaked at 105--110 GeV, in agreement with 
ref.\cite{aleph4j}. The slight downward shift of the peak from 
$2 m_{\Ch}$ is due to the suppression of the soft jets ($E < \sigma_E$)
as discussed above. On the other hand, the four-body decay of stop via
$\Ne$ (\ref{st_cn} \& \ref{neut_decay}) is seen to result in a very 
broad distribution. Finally, it should be noted that the normalization
of the solid curves in Figs.\ref{fig:mass_distr}({\em a,b}) 
are about twice as large as the ALEPH event size ($\sim 9$). One could 
reduce our event rate by assuming a smaller sneutrino 
mass ($\sim 150 \gev$). On the other hand, several of the ALEPH cuts 
such as the number of charged tracks and individual jet masses etc., could
not be incorporated into our parton level Monte Carlo. As these will,
typically, result in a loss of efficiency, we leave this excess in 
normalization.

In summary, relatively light stops ($m_{\tilde t_1} \lsim 60 \gev$) and 
charginos offer a viable MSSM solution to the $R_b$ anomaly within the 
constraints of top quark decay and gaugino mass unification. Assuming
an $R$-parity violating Yukawa coupling $\lambda''$ in the superpotential, 
it is possible to bring down the chargino mass below 60 GeV as well, while
respecting the $Z$-decay constraint. Consequently, the pair production 
of stops via chargino decay at LEP1.5 can offer a viable solution to the 
ALEPH 4-jet anomaly as well. The direct decay of stop into a light quark 
pair is expected to be the dominant mode for 
$\lambda''_{tds} \gsim 5 \times 10^{-5}$. 
This can account for the essential features of the ALEPH events at a
quantitative level.

We gratefully acknowledge discussions with Sunanda Banerjee, 
Piotr Chankowski, Atul Gurtu and Stefan Pokorski, and computational help
from Sreerup Raychaudhuri.

\end{document}